\documentclass{article}
\usepackage{spconf,amsmath,graphicx}
\usepackage{cite}
\usepackage{spconf,amsmath,graphicx}
\usepackage{amssymb}
\usepackage{mathrsfs}
\usepackage{amsmath}
\usepackage{comment} 
\usepackage{here}
\usepackage{bm}
\usepackage{url}
\usepackage{booktabs}
\usepackage{enumitem}

\DeclareMathOperator{\multihead}{Multihead}

\DeclareMathOperator{\head}{head}
\DeclareMathOperator{\softmax}{Softmax}

\DeclareMathOperator{\concat}{Concat}

\DeclareMathOperator{\onehot}{Onehot}
\title{Multi-view and Multi-modal Event Detection\\ Utilizing Transformer-based Multi-sensor fusion}
\vspace{-15pt}
\name{
Masahiro Yasuda$^{\dagger}$,
Yasunori Ohishi$^{\dagger}$,
Shoichiro Saito$^{\dagger}$,
Noboru Harada$^{\dagger}$,
}
\vspace{-16pt}
\address{$^\dagger$NTT Corporation, Japan
\vspace{-13pt}
}

\begin{document}
\ninept
\maketitle

\begin{abstract}
\vspace{-3pt}
We tackle a challenging task: multi-view and multi-modal event detection that detects events in a wide-range real environment by utilizing data from distributed cameras and microphones and their weak labels.
In this task, distributed sensors are utilized complementarily to capture events that are difficult to capture with a single sensor, such as a series of actions of people moving in an intricate room, or communication between people located far apart in a room. For sensors to cooperate effectively in such a situation, the system should be able to exchange information among sensors and combines information that is useful for identifying events in a complementary manner.
For such a mechanism, we propose a Transformer-based multi-sensor fusion (MultiTrans) which combines multi-sensor data on the basis of the relationships between features of different viewpoints and modalities.
In the experiments using a dataset~\footnote{\scriptsize{Our dataset ``MM-Office'' is publicity available at \url{https://github.com/nttrd-mdlab/mm-office}} and \url{https://doi.org/hg4s}} newly collected for this task, our proposed method using MultiTrans improved the event detection performance and outperformed comparatives.
\end{abstract}

\begin{keywords}
multi-view, cross-modal, event detection, distributed sensor, weakly-supervised learning
\end{keywords}

\vspace{-5pt}
\section{Introduction}
\vspace{-3pt}
\label{sec:intro}
Our goal is to develop a system that supports daily life via recognizing and understanding a wide-range of human activities utilizing linked multiple sensors. % that already exist in our surroundings, such as surveillance cameras and smartphones. 
Examples of such systems include a wide-area security system that detects people behaving suspiciously and signs of criminal activity, an auto-house-cleaning system that disinfects areas touched by people to prevent the spread of infectious diseases, and an automatic work support system for employees in offices. 
As the first step towards such applications, this study focuses on understanding human activity in a real office environment.

Event detection has been studied as a fundamental technology for such applications.
Human action detection is a vital study topic in computer vision. High performance is achieved by utilizing a deep neural network (DNN) to extract and classify features~\cite{actionrec1,actionrec2}.
Sound event detection (SED) is a similar task but it uses sound as a modality~\cite{sed1,sed2}.
However, event detection techniques using a single or limited number of sensors with a single modality are not sufficient to understand human activity in a wide-range real environment. 

\begin{figure}[tb!]
  \begin{center}
  \includegraphics[width=0.7\linewidth]{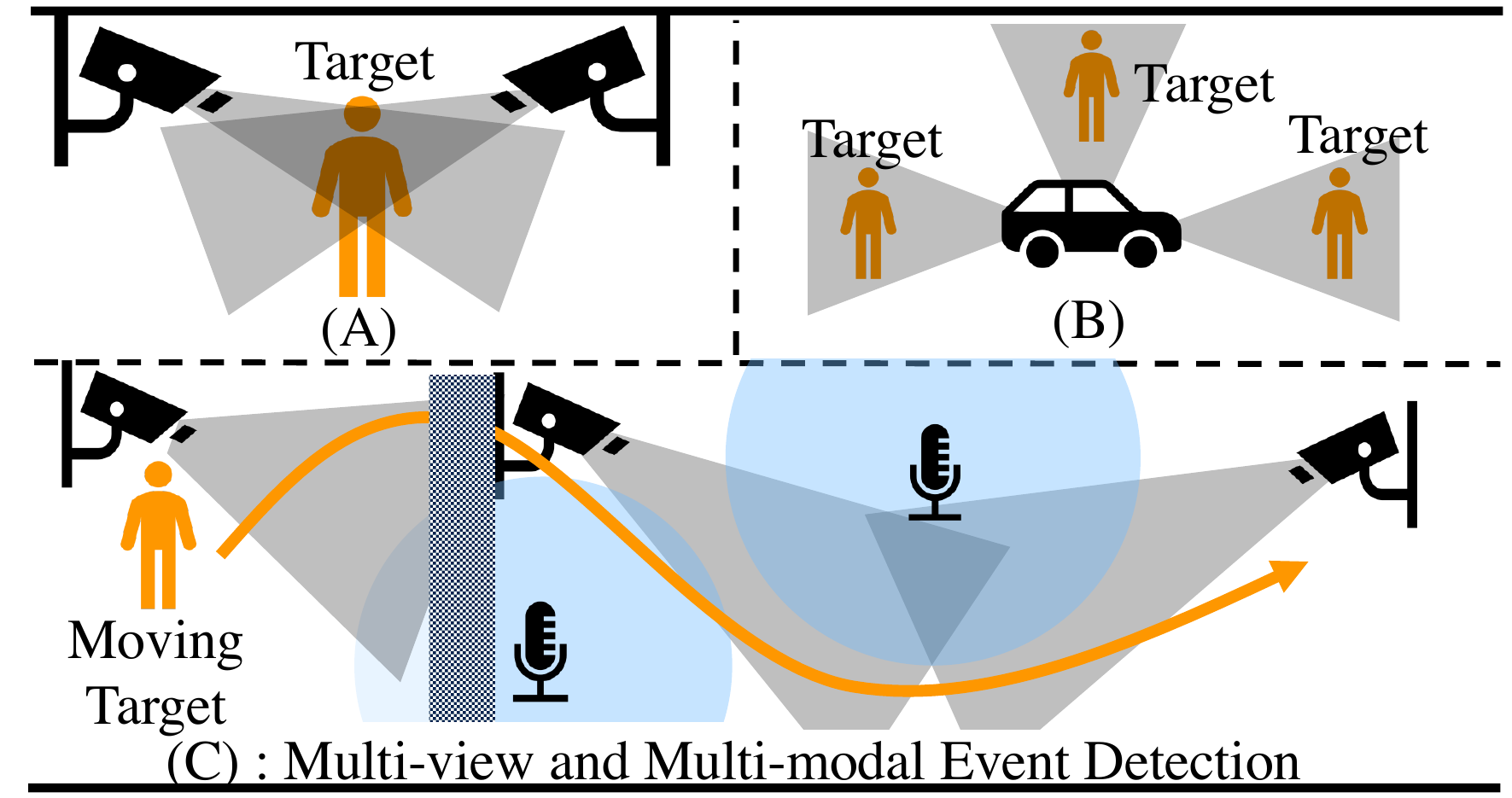}
\vspace{-10pt}
\caption{Possible configurations of multiple sensor arrangement.}
  \label{fig:multiview}
  \end{center}
  \vspace{-26pt}
 \end{figure}

%Multiview
Several existing works have involved multiple sensors. 
For example, multiple cameras are arranged so as to surround a target existing in a narrow region, as illustrated in Fig.~\ref{fig:multiview} (A)~\cite{mvcnn,danet,mmact,mva5}. 
Another example utilizing multiple sensors is a self-propelled machine as illustrated in Fig.~\ref{fig:multiview} (B)~\cite{crossview,deepfusion}. 
In terms of sensor arrangement, these two examples are not appropriate for understanding a wide-range real environment,
because the former is aimed at capturing a narrow region more correctly,
whereas the latter is aimed at capturing the first-person environment. For covering wide-range areas, a straightforward approach is to use distributed sensors as illustrated in Figure\ref{fig:multiview} (C).
In fact, recent studies have shown that the spatial features acquired by distributed microphones help to understand acoustic scenes~\cite{distmic}.
%Multimodal
In addition, the combination of different modality sensors is also expected to be useful for identifying events that cannot be distinguished by a single modality~\cite{multi1,multi2}.  

Therefore, we tackle a challenging task, multi-view and multi-modal event detection, for the purpose of understanding human activities in a wide-range real environment.
The multi-view and multi-modal event detection involves identifying the class and onset/offset time of an event utilizing distributed cameras and microphones.
Although typical event detection algorithms utilize strong labels that indicate both the class and the onset/offset time, in our task, the annotation cost of strong labels is huge because there are multiple modalities and viewpoints.
For this practical constraint, we consider our task under a condition in which only weak labels indicating classes of the events included in a single clip without temporal information are given.

The key to successful multi-view and multi-modal event detection is inter-sensor cooperation, that is, acquiring a good joint representation combining sensor data that complement each other and hold sufficient information about the event.
In contrast to conventional event detection and action recognition tasks that utilize multiple sensors, only a limited number of sensors clearly capture the target, as we can see in Fig.~\ref{fig:multiview} (C). In this condition, several sensor data are uninformative, and such redundant data could become noise in the detection of the target event. In addition, previous studies reported that the effective modality or viewpoints are different depending on the class of event~\cite{mva4,multi1}. Considering the above, multi-view and multi-modal event detection requires a mechanism to pay attention to appropriate sensors depending on the situation: when, where, and what event happened.

\begin{figure*}[tb!]
  \begin{center}
  \includegraphics[width=0.89\linewidth]{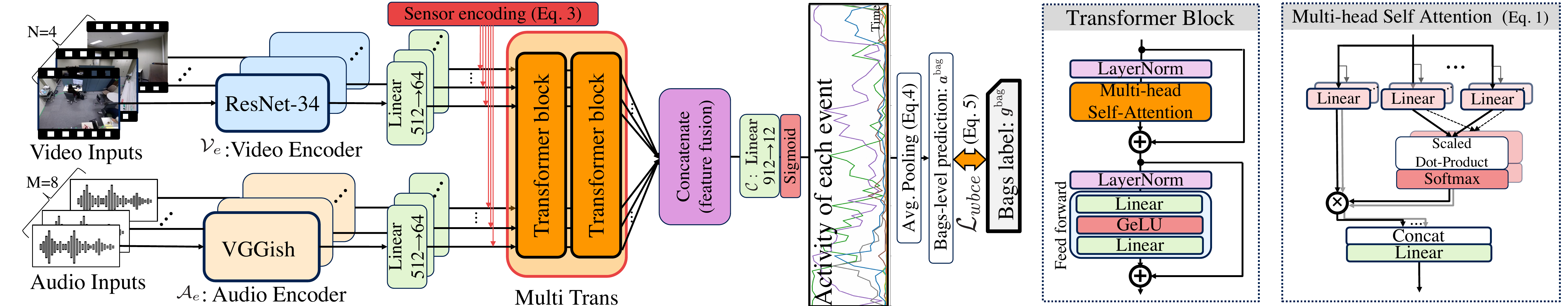}
  \vspace{-10pt}
  \caption{Network architecture of proposed method.}
  \label{fig:architecture}
  \end{center}
  \vspace{-25pt}
 \end{figure*}

As such a mechanism, we propose Transformer-based multi-sensor fusion (MultiTrans), which combines sensor data useful for identifying events based on inter-sensor relationship.
Unlike a conventional Transformer that takes as input a sequence of words, pixels or audio signals~\cite{attention,vit,asrtrans}, MultiTrans takes as input a sequence of embeddings obtained from each sensor, enabling it to handle a large number of sensors and extract a joint representation.
The self-attention in a Transformer is expected to focus on features that are useful for the task, as evidenced by various research areas~\cite{attention,vit,asrtrans}.
Therefore, MultiTrans is expected to be suitable as a situation-aware system as required for multi-view and multi-modal event detection.

To verify the effectiveness of MultiTrans for the multi-view and multi-modal event detection task, we newly collected a multi-view and multi-modal dataset in an office environment (MM-Office).
Unlike existing multi-view datasets~\cite{mvdataA1,mvdataA2,mmact,mvdataB1, mvdataB2,mvdataB3}, MM-Office aims to capture human activities across the entire office, utilizing cameras and microphones distributed across the room.  In a validation experiment using MM-Office, our proposed method using MultiTrans outperformed comparison methods that used the existing multi-sensor fusion method in the multi-view and multi-modal event detection task.
\vspace{-14pt}
\section{Related Studies}
\vspace{-4pt}
\label{sec:RelatedStudy}
Multi-sensor fusion refers to the fusion of sensor data or their features from multiple viewpoints and modalities for various tasks. It includes multi-modal fusion, multi-view fusion, and both of them. 

Various multi-modal fusion methods have been proposed for extracting inter-modal relationship and fusing representations~\cite{multimodalfusion1,multimodalfusion2,multimodalfusion3,transfuser}. In particular, the use of multi-head self-attention (MHSA) based frameworks such as Transformer has recently been reported to be effective for linking different modalities~\cite{videobert,transfuser}. 
MHSA is defined as the case where $Q=K=V$ in the following equation:
\vspace{-3pt}
\begin{equation}
\begin{split}
\multihead(Q, K, V) = \concat(\head_1,\ldots,\head_h)W_O \\
{\rm where}\hspace{2pt}\head_i = \softmax\left(\frac{QW^Q_i(KW^K_i)^{tr}}{\sqrt{d_k}}\right)VW^V_i
\end{split}
\vspace{-7pt}
\end{equation}
Here, $W^Q_i$, $W^K_i$, and $W^V_i$ are projection parameters, and $d_k$ are dimension of $K$. By applying different projections on the input for each head, it is possible to model the relationship between the input sequences from multiple perspectives. Although MHSA was originally proposed for natural language processing, now it has been reported that by using joint sequences from two modalities as input (e.g., video and word or image and light detection and ranging (LiDAR) image), it is possible to extract the relationship between them~\cite{videobert,transfuser}.

On the other hand, there are few studies on multi-view joint representation, and to the best of our knowledge, only Wang {\it et al.}~\cite{danet} have explicitly considered the inter-view relationship. In their work, a conditional random field (CRF) is introduced to exchange messages among views~\cite{crfpaper}. Considering such an inter-views relationship enables multiple views to cooperate more effectively.

\vspace{-5pt}
\section{Proposed method}
\label{sec:proposedMetod}
\vspace{-3pt}
%In this section, we describe the proposed method for multi-view cross-modal event detection.
\subsection{Problem settings}
\vspace{-3pt}
To explain the proposed method, we clarify the problem setting and notation of multi-view and multi-modal event detection.
Here, we consider a system that takes input sensor data from $M$-channel microphones and $N$-channel cameras that are distributed in the given environment.
The input sequences for this task are the acoustic features obtained from the microphones and the video features obtained from the cameras.
These input sequences are denoted as $\Psi=(\psi_1,\ldots,\psi_T)$, where $\psi_{\tau}=(\phi_{1,\tau},\ldots,\phi_{S,\tau})^{tr}$ is the input features at time index $\tau\in\{1,\ldots,T\}$, $\phi_{s,\tau}\in\mathbb{R}^{D}$ is an input feature at sensor index $s\in\{1,\ldots,S=M+N\}$. 
The outputs of this task are the activation of event $\bm{A} = (\bm{a}_1,\bm{a}_2,...,\bm{a}_T)$, where $\bm{a}_{\tau}=(a_{1,\tau},\ldots,a_{C,\tau})^{tr}\in \{0,1\}^{C}$ is the indicator of event at time index $\tau$, and $C$ is the number of event classes.
Let ground truth correspond to activity of event $\bm{A}$ as $\bm{G}=(\bm{g}_1,\ldots,\bm{g}_T)$.
In weakly-supervised setting, such an frame level ground truth is not available.
Instead, a ground truth called bags label $\bm{g}^{\mbox{\scriptsize bag}}=(g^{\mbox{\scriptsize bag}}_1,\ldots,g^{\mbox{\scriptsize bag}}_C)\in\{0,1\}^C$ is given, defined as follows:
\vspace{-5pt}
\begin{equation}
    g^{\mbox{\scriptsize bag}}_c =   
    \begin{cases}
    1\hspace{10pt}{\rm if}\hspace{10pt}^{\exists}g_{c,\tau}\,:\,g_{c,\tau}=1;\\
    0\hspace{10pt}{\rm if}\hspace{10pt}^{\forall}g_{c,\tau}\,:\,g_{c,\tau}=0.
    \end{cases}
\end{equation}

\vspace{-14pt}
\subsection{Basic concept}
\vspace{-2pt}
Our primary interest is to extract a good joint representation harmonizing features of multiple sensors. 
A good joint representation for multi-view and multi-modal event detection should help to extract appropriate information to identify the event.
However, in a real environment, observations from not all sensors (i.e., all viewpoints and modalities) satisfy this condition. 
In the visual modality, events hidden by occlusion cannot be observed, and distant events are difficult to follow in detail.
In audio modalities, noise and distance attenuation may cause some sensors to be uninformative.
Besides, which viewpoints and modalities are valid depends on what event occurs. 
For example, although audio modality is obviously suitable for capturing human voices, it is better to refer to video modality to determine whether that voice comes from a phone conversation or a face-to-face conversation.
Existing studies of multi-view and multi-modal event detection have not taken such limitations into account since these deal with cases where all sensors are clearly useful for identifying events~\cite{mmact,mva5}.
From the above, successful multi-sensor fusion for multi-view and multi-modal event detection requires combining essential sensors data depending on the situation: when, where and what event happened.

Therefore, we propose MultiTrans to associate and combine all features obtained from multiple sensors in accordance with the situation.
MultiTrans is implemented as a Transformer that takes as input a sequence of embeddings extracted from each sensor via an encoder such as a convolutional neural network (CNN), and is expected to model the inter-sensor relationship just as the original Transformer modeled the inter-word relationship.

%%%=====

Here, since Transformer does not distinguish the order of the input sequences, the sensor index information to input embeddings needs to be provided in advance.
For this purpose, the original Transformer uses positional encoding~\cite{attention}, but since our sensor sequences do not have a continuous relationship between adjacent inputs like word sequences, discrete encoding is considered more appropriate. For this reason, we introduce the following sensor encoding:
\vspace{-2pt}
\begin{equation}
    \tilde{\phi}_{s,\tau}= \concat(\phi_{s,\tau}, \onehot_S(s)) \in \mathbb{R}^{D+S},
    \label{eq:viewenc}
\end{equation}
where $\phi_s\in\mathbb{R}^{D}$ is the feature from the $s-$th sensor, and $\onehot_S(s)$ is the one-hot vector in which the $s-$th element is equal to 1. 

\vspace{-7pt}
\subsection{Implementation details}
\vspace{-2pt}
Fig.~\ref{fig:architecture} shows the network architecture of our proposed method.
The input of the system is synchronized $M$-channel videos and $N$-channel audio signals.
As in other existing works using multiple sensors~\cite{mvcnn,danet,mmact,crossview}, the shared video encoder $\mathcal{V}_e$ and shared audio encoder $\mathcal{A}_e$ are used to extract embeddings. The video encoder $\mathcal{V}_e$ is implemented as ResNet-34~\cite{resnet34} pre-trained on ImageNet, minus the output layer. 
In the audio encoder $\mathcal{A}_e$, the input audio is firstly transformed into the log-absolute value of the short-time Fourier transform (STFT) spectrogram, and then a Mel-filter bank is applied. The audio embeddings are extracted using VGGish~\cite{vggish} pre-trained with AudioSet. The extracted audio and video features are embedded in the $D$-dimensional vectors with a linear layer. After applying the sensor-encoding Eq. (\ref{eq:viewenc}), these audio and video embeddings are inputted to MultiTrans. MultiTrans consists of stacked Transformer blocks, and MHSA of each block has $H$ heads.
The classifier $C$ for estimating event activity $\bm{\tilde{a}}_{\tau}$ from the fused feature is implemented as a linear layer with a sigmoid activation function.
%The output of the system, activation of event $\bm{a}_{\tau}$, is obtained by applying a hard threshold to $\bm{\tilde{a}}_{\tau}$.
The activation of the event $\bm{a}_{\tau}$, the output of the system, is 1 if $\bm{\tilde{a}}_{\tau}$ exceeds a fixed threshold and 0 otherwise.
\begin{figure}[tb!]
  \begin{center}
  \includegraphics[width=0.75\linewidth]{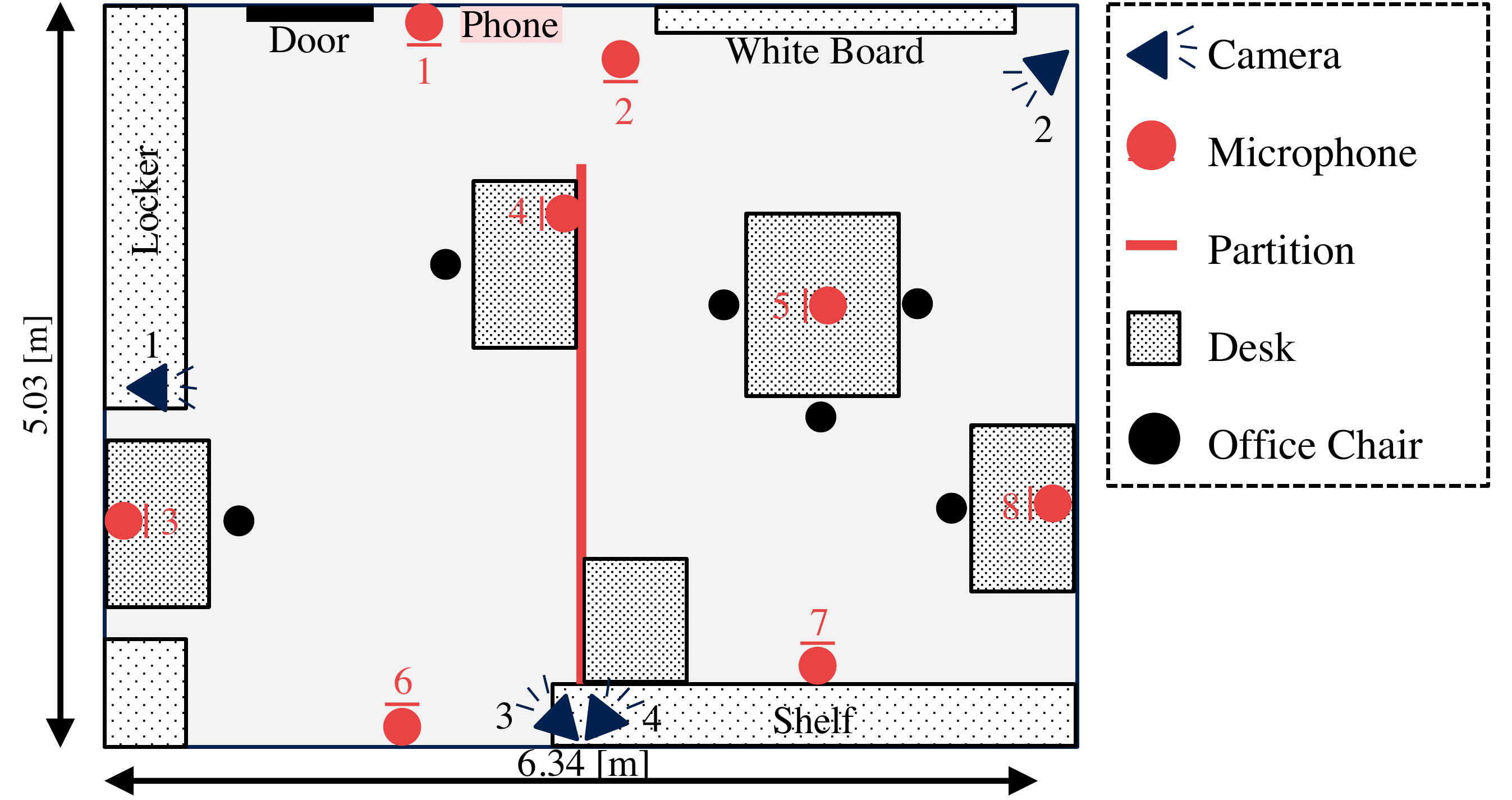}
  \vspace{-12pt}
  \caption{Room and sensors setup of MM-Office dataset}
  \label{fig:roomsetup}
  \end{center}
  \vspace{-23pt}
 \end{figure}
The whole system is trained using the weak label $\bm{g}^{\mbox{\scriptsize bag}}$ in a multiple-instance learning (MIL) manner~\cite{mil}, often used in weakly-supervised sound event detection~\cite{sed3}.
In the MIL scheme, the following bags-level prediction $\bm{\tilde{a}}^{\mbox{\scriptsize bag}}$ is first calculated from the obtained event activity sequence as:
\vspace{-10pt}
\begin{equation}
\label{eq:bagpred}
%\hspace{-2mm}
\bm{\tilde{a}}^{\mbox{\scriptsize bag}} = \frac{1}{T}\sum_{\tau=1}^{T}\bm{\tilde{a}}_{\tau}
\vspace{-3pt}
\end{equation}
Since the dataset we used has a large class imbalance, as shown in the Table\ref{tb:eventlist}, we use the following weighted BCE~\cite{wbce}:
\vspace{-5pt}
\begin{equation}
\thickmuskip=0mu
\medmuskip=0mu
\thinmuskip=0mu
\mathcal{L}=-\frac{1}{B}\sum_{b=1}^{B}\sum_{c=1}^{C}
w_c\left(g^{\mbox{\scriptsize bag}}_{c,b}\log(\tilde{a}^{\mbox{\scriptsize bag}}_{c,b}) \right.
\left. +(1-g^{\mbox{\scriptsize bag}}_{c,b})\log(1-\tilde{a}^{\mbox{\scriptsize bag}}_{c,b}) \right)
\vspace{-5pt}
\end{equation}
where $w_c$ is the reciprocal of the total number of events in dataset belonging to the $c-$th class, $b\in\{1,\ldots,B\}$ is the index of batch in mini-batch training.
\vspace{-5pt}
\section{Experiments}
\vspace{-3pt}

\subsection{Experimental Setup}
\vspace{-2pt}
This section describes the setup of verification experiments to evaluate the effectiveness of the proposed method.

\vspace{3pt}
\noindent
{\bf Dataset: }We collected a new dataset, called the multi-view and multi-modal office dataset (MM-Office), which fits our problem setting. The recording environment is an office room as shown in Fig.~\ref{fig:roomsetup}
, and a partition acting as an occlusion is installed in the center of the room. In this room, 1 to 3 people act in accordance with 11 scenes assuming the daily work.
Each scene contains 12 classes of events shown in Table~\ref{tb:eventlist}.
These events are recorded simultaneously using eight non-directional microphones and four cameras. 
This room and sensor setting is a somewhat realistic setup for applications such as understanding or logging the behavior of employees in a specific office. 
The audio and video clips are divided into scenes, each of which is about 30 to 90 seconds. The amount of data was 880 clips per point and sensor, of which 704 were used for learning and 166 for evaluation.
The labels available for training are given as multi-labels that indicate which each clip contains what event. To use for evaluation, only the test data is annotated with a strong label containing the onset/offset time of each event in accordance with the event definition as shown in Table~\ref{tb:eventlist}.

\begin{table}[t!]
\vspace{-10pt}
\centering
\caption{Event list in MM-Office dataset}
\label{tb:eventlist}
\scalebox{0.8}[0.8]{
\begin{tabular}{l|ll}
\toprule
Class&\# event&description of each event\\ \midrule
eat&90&Eating lunch in the desk. \\
tele&164&Talking to someone remotely using a telephone. \\
chat&412&Talking directly with people in the room. \\
meeting&90&Gathering at a large desk for a meeting. \\
takeout&123&Take out something from shelf or box. \\
prepare&84&Preparing equipment for a teleconference. \\
handout&162&Handing out food or packages. \\
enter&225&Opening the door to enter a room and greeting. \\
exit&90&Opening the door and leaving the room. \\
stand up&195&Stand up from a chair.\\
sit down&372&Sitting down on a chair. \\
phone&60&Ringing a phone or cell phone. \\ \midrule
b.g.&-&Unoccupied rooms or mere desk work or walking.\\ \bottomrule
\end{tabular}
}
\vspace{-18pt}
\end{table}

\begin{table*}[htbp]
\centering
\caption{multi-view and multi-modal event detection performances.}
\label{tb:resultdet}
\scalebox{0.8}[0.8]{
\begin{tabular}{l|cccccccccccc|c}
\toprule
Model& eat&tele &chat& meeting & takeout
& prepare & handout & enter & exit & stand up & sit down & phone & mAP\\\midrule
(A-1) Sum&89.8\%&60.3\%&44.2\%&86.6\%&63.1\%&52.0\%&6.8\%&29.6\%&30.5\%&4.7\%&4.9\%&0.7\%&39.4\%\\
(A-2) Max&{\bf 92.6}\%&{\bf 68.5}\%&46.1\%&85.1\%&68.5\%&31.6\%&7.5\%&46.5\%&24.6\%&5.5\%&6.2\%&{\bf 7.8}\%&40.9\%\\
(A-3) Concat&85.3\%&61.8\%&47.6\%&81.0\%&68.4\%&45.5\%&5.5\%&31.5\%&29.3\%&6.9\%&4.7\%&0.6\%&39.0\%\\ \midrule
(B-1) CRF-Sum&66.2\%&56.2\%&56.1\%&{\bf 91.0}\%&68.3\%&37.6\%&8.4\%&31.8\%&28.6\%&4.7\%&4.3\%&0.6\%&37.8\%\\
(B-2) CRF-Max&83.6\%&67.6\%&52.5\%&86.1\%&61.2\%&42.0\%&4.5\%&{\bf 53.5}\%&22.2\%&{\bf 10.3}\%&4.8\%&2.6\%&40.9\%\\
(B-3) CRF-Concat&83.7\%&60.2\%&46.7\%&83.4\%&{\bf 78.5}\%&47.2\%&6.0\%&50.4\%&39.0\%&3.9\%&5.4\%&7.2\%&42.6\%\\ \midrule
(C-1) MultiTrans-Sum&86.3\%&55.9\%&52.7\%&74.9\%&76.1\%&54.4\%&6.2\%&45.3\%&28.5\%&5.5\%&5.8\%&0.9\%&41.0\%\\
(C-2) MultiTrans-Max&84.0\%&50.9\%&{\bf 73.7}\%&83.9\%&75.4\%&50.5\%&7.9\%&34.2\%&{\bf 40.1}\%&5.3\%&5.3\%&1.2\%&42.7\%\\
(C-3) MultiTrans-Concat&59.9\%&62.4\%&73.6\%&88.5\%&70.6\%&{\bf 68.2}\%&{\bf 14.2}\%&42.9\%&31.2\%&6.0\%&{\bf 7.2}\%&4.9\%&{\bf 44.1}\%\\ \bottomrule
\end{tabular}
}
\vspace{-11pt}
\end{table*}

\vspace{3pt}
\noindent
{\bf Hyper parameters: } 
The input of the system is the audio signal obtained from the microphone with $M=8$ channels and the video signal obtained from the camera with $N=4$ channels. 
We fix the input clip length at $T=25.6 $sec. Since the length of each training data is about 30 to 90 seconds and longer than $T$, clips of length $T$ are randomly sampled from the training data at every iteration.
The video input downsampled from 30 to 2.5 fps and compressed the resolution to 112 $\times$112.
The sampling frequency of audio input was downsampled from 48kHz to 16 kHz, and the STFT spectrogram was extracted using a 400 point Hanning window with a 160 point shift.
The dimension of the Mel-filter bank was 64. The resulting Mel-spectrogram is then divided into 32 segments of 80 time frame lengths to align it with the frame rate of the video inputs.
The output dimensions of the audio and video encoder $\mathcal{A}_e$ and $ \mathcal{V}_e$ are both 512, and the embedding dimension is $D = 64$. 
In MultiTrans, the Transformer block was stacked two layers, and the MHSA in the Transformer block had $H=4$ number of the heads. In the preliminary experiments, the performance was degraded at a lower or higher number of heads $H$.
%The parameters of audio encoder $\mathcal {A}_e $ were fixed except for CNN of the last layer. For the video encoder $ \mathcal{V}_e $, all parameters were retrained.
The AdamW optimizer was used for all training, the initial learning rate was set at 0.01 and exponentially decayed to 0.1 times at the end epoch, weight decay paramete was set to 0.01~\cite{adamw}. The maximum epoch of learning was fixed at 50.

\vspace{3pt}
\noindent
{\bf Comparison methods and evaluation metrics: }
%To evaluate the effectiveness of the proposed method for multi-view and multi-modal event detection in terms of modeling inter-sensor relationships, we compare it with a baseline method that does not utilize inter-sensor modeling and a comparison method utilizing CRF~\cite{danet}. 
To evaluate the effectiveness of the proposed method (denoted as {\bf (C)} ) for multi-view and multi-modal event detection in terms of modeling inter-sensor relationships, we compare it with two comparison methods; {\bf (A)} A baseline method that simply excludes the MultiTrans from the proposed architecture; {\bf (B)} A CRF-based method that replacing MultiTrans by CRF[6] in the proposed architecture.
To the best of our knowledge, the CRF-based method is the only existing work that explicitly considers the inter-view relationship.
CRF and MultiTrans output the same number of features as the number of sensors, i.e., there is a choice of how to fuse these features further.
The original network using CRF, Dividing and Aggregating Network (DA-Net), uses a method called score fusion, but this is premised on video-only input and is not suitable for our problem setting. 
In this experiment, we investigated the combination of the comparative method and the proposed method with the three standard fusion methods: {\bf (1) Sum}: sum of embeddings; {\bf (2) Max}: max pooling of the embeddings along sensor dimension; {\bf (3) Concat}: concatenation of embeddings along feature dimension that adopted in the proposed method. In the results section, combinations of the inter-sensor relationship modeling methods (A)-(C) and the fusion methods (1)-(3) are denoted by a combination of letters and numbers (e.g., (A-1)).
% In this experiment, the comparison method and the proposed method are combined with three standard fusion methods: {\bf Sum}: sum of embeddings; {\bf Max}: max pooling of the embeddings along sensor dimension; {\bf Concat}: concatenation of embeddings along feature dimension.
As evaluation metrics, we adopt mean average precision (mAP), which are widely accepted in detection tasks. The mAP is the average of the precision at different recalls and does not depend on the threshold.

\begin{figure}[t!]
  \centering
  \includegraphics[width=0.96\linewidth]{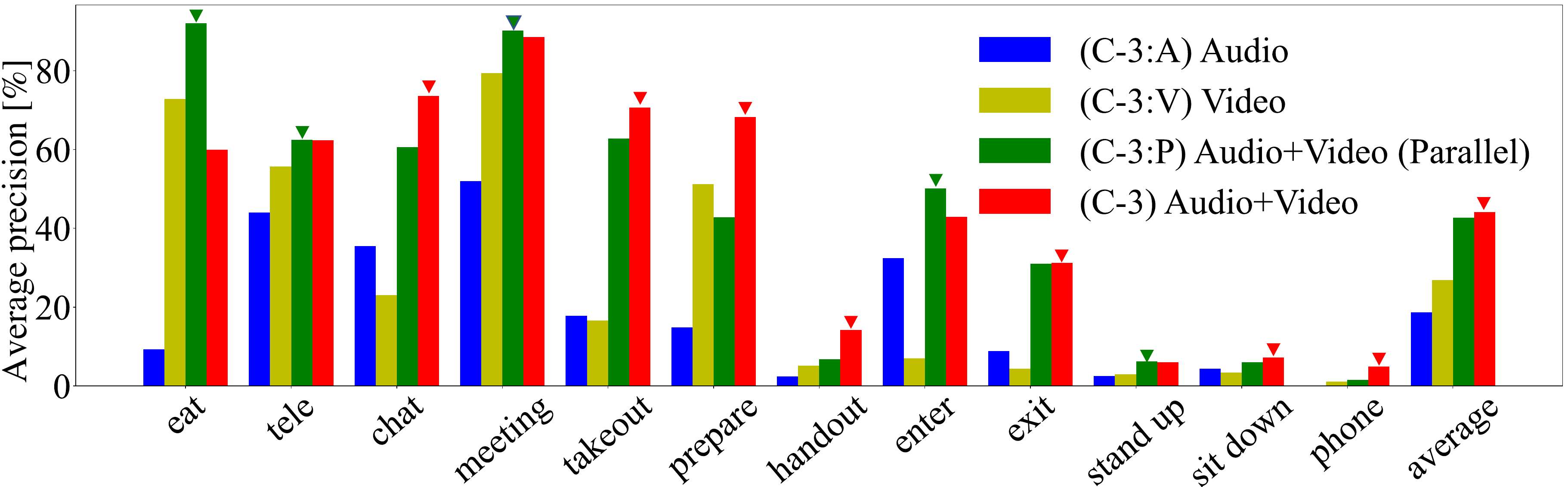}
      \vspace{-13pt}
 \caption{Comparison of audio and video modality in AP scores.}
  \label{fig:audiovisual}
  \vspace{-16pt}
\end{figure}

\vspace{-8pt}
\subsection{Results}
\vspace{-4pt}
Table~\ref{tb:resultdet} shows the event detection performance of the proposed method and the comparison methods. First, the proposed (C-3) outperforms the other methods, especially in the ``prepare'', ``chat'', and ``handout'' classes. Identifying these classes requires coordination among sensors. In fact, ``prepare'' is a complex and extensive event as taking a device for teleconference from the shelf, carrying it to the table, and setting it up. Next, comparing (A-1) to (C-1), (A-2) to (C-2), and (A-3) to (C-3) respectively, it is observed that MultiTrans improves the performance of the system regardless of whether Sum, Max, or Concat is used. 
Note that, considering the variation in performance between (C-1) to (C-3), there may be room for performance improvement in future work by further combining more sophisticated feature fusion with MultiTrans output. These results suggest that the introduction of MultiTrans effectively performs multi-view multimodal event detection, especially in identifying behaviors that span a wide-range. 

However, in all methods, the low mAP is obtained in ``sit down,'' ``stand up,'' and ``phone,''. This is because the time duration of these classes is short, making it difficult to identify the onset/offset time in a weakly-supervised setting. Further improvement of the performance of such a short event in the untrimmed video is a future task.

As an ablation study of the proposed (C-3) on the use of multiple modalities, we investigated the following cases in (C-3:A) audio-only, (C-3:V) video only, (C-3:P) both audio and video input, but using MultiTrans independently for audio and video. Fig.~\ref{fig:audiovisual} shows the comparison between the proposed method and each ablation method. First, either of the methods with both audio and video overcome uni-modal methods in all classes and overall performance. Next, (C-3) outperforms (C-3:P) especially in ``chat'', ``takeout'', and ``prepare'' classes. The inter-modality relationship between audio and video is considered to be important for the identification of these classes. 
%For example, in order to identify the ``chat'' class, it is necessary to combine video information to distinguish it from other events containing human voices such as ``tele'' and ``meeting''. 
For example, in ``take out'', the precision in audio-visual is quite higher than the accuracy in single modality. This suggests that the action of taking out an object from a shelf may be distinguished from other events only when sound and video are combined. 
These results show that the use of multiple modalities and the modeling of inter-modality relationships by MultiTrans both contribute to performance improvement.

\begin{figure}[t!]
  \centering
    \vspace{-5pt}
  \includegraphics[width=\linewidth]{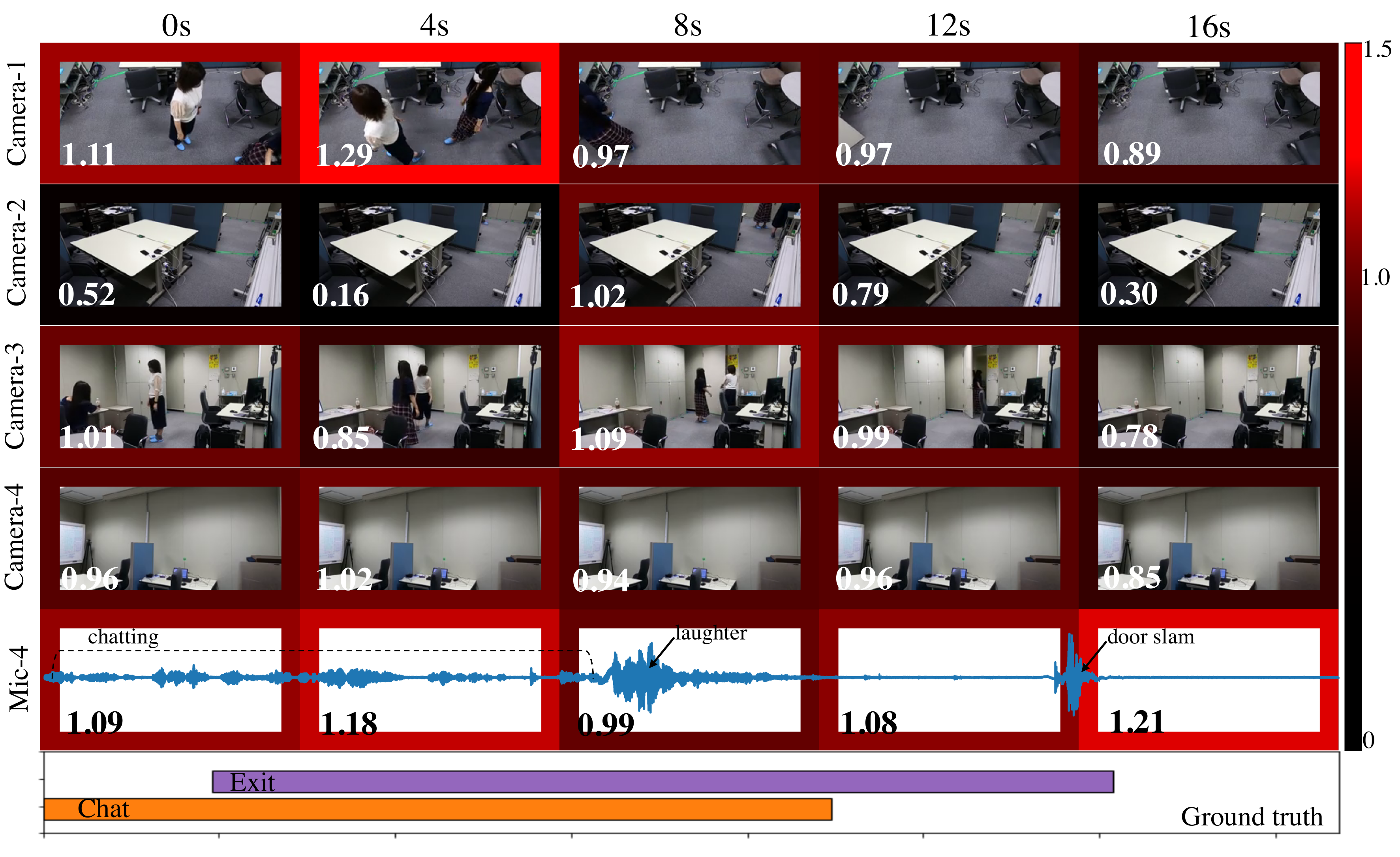}
    \vspace{-21pt}
    \caption{Visualization of attention weights of multi-head self-attention module in MultiTrans.}
  \label{fig:attention}
  \vspace{-13pt}
\end{figure}

To investigate the property of MultiTrans, we focused on the attention weights of one head in the last MHSA layer and visualized them. Fig.~\ref{fig:attention} shows an example of a visualization of the scene in which one person calls out to another person, who is sitting, and then exits the room from the door. In this visualization, the attention weight is represented as the color of the frame of each image, and is normalized to be 1.0 when all sensors are given equal attention. First, the most outstanding attention is observed on Camera-1 at 4s. This is a reasonable choice because it is the viewpoint that best captures two people talking as they walk towards the exit. Another noteworthy point is the increase in the attention weight of the microphone in 16s. This suggests that the door opening and closing sound is important information in identifying ``exit.''

\vspace{-6pt}
\section{Conclusion}
\vspace{-3pt}
In this study, we proposed Transformer-based multi-sensor fusion (MultiTrans), a mechanism to combine features of effective modalities and viewpoints for a multi-view and multi-modal event detection task in a real office environment.
%The input of MultiTrans is a sequence of embeddings from each sensor. 
MultiTrans is expected to combine sensor data complementary based on inter-sensor relationships, just as the original Transformer could model inter-word relationships well.
In a validation experiment for the dataset recorded in a real office environment, the proposed method using MultiTrans 
outperforms a standard fusion method that does not consider inter-sensor relationships, and a comparative method that introduces conditional random fields (CRFs) to model the inter-sensor relationships.
Therefore, we conclude that MultiTrans is effective for multi-view and multi-modal event detection.

\newpage

\bibliographystyle{IEEEbib}
\footnotesize
\bibliography{refs}

\end{document}